\documentclass[apjl]{emulateapj}

\usepackage{natbib} 
\usepackage{graphicx}
\usepackage{threeparttable}
\usepackage{amsmath}
\usepackage{color}
\usepackage{booktabs}
\usepackage{multirow} 
\usepackage{etex,xy} 
\xyoption{all} 

\definecolor{orange}{rgb}{0.93, 0.57, 0.13}
\definecolor{green}{rgb}{0.01, 0.75, 0.24}
\definecolor{gray}{rgb}{0.66,0.66,0.66}

\UseCrayolaColors

\newcommand{\GILDAS}{\texttt{GILDAS}}
\newcommand{\CLASS}{\texttt{CLASS}}
\newcommand{\CLIC}{\texttt{CLIC}}
\newcommand{\MAPPING}{\texttt{MAPPING}}

\newcommand{\ie}{\emph{i.e.}}
\newcommand{\eg}{e.g.}
\newcommand{\emm}[1]{\ensuremath{#1}}   
\newcommand{\emr}[1]{\emm{\mathrm{#1}}} 
\newcommand{\chem}[1]{\ensuremath{\mathrm{#1}}}
\newcommand{\Tkin}{\emm{T_\emr{kin}}}

\newcommand{\nH}{\emm{n_\emr{H}}}
\newcommand{\NH}{\emm{N_\emr{H}}}

\newcommand{\hh}{\emr{H_2}}                  
\newcommand{\h}{\emr{H}}                  
\newcommand{\cch}{\emr{C_2H}}                  
\newcommand{\ccch}{\emr{C_3H}}                  
\newcommand{\Cccch}{$c$-\emr{C_3H}}                  
\newcommand{\Lccch}{$l$-\emr{C_3H}}                  
\newcommand{\ccchh}{\emr{C_3H_2}}                  
\newcommand{\Cccchh}{$c$-\emr{C_3H_2}}                  
\newcommand{\Lccchh}{$l$-\emr{C_3H_2}}                  
\newcommand{\cccch}{\emr{C_4H}}                  
\newcommand{\ccchp}{\emr{C_3H^+}}                  
\newcommand{\lccchp}{$l$-\emr{C_3H^+}}                  
\newcommand{\cp}{\emr{C^+}}                  
\newcommand{\dcop}{\emr{DCO^+}}                  

\renewcommand{\deg}{\emm{^\circ}}
\newcommand{\pccm}{~\rm{cm}^{-3}}

\newcommand{\kms}{\emr{\,km\,s^{-1}}}

\newcommand{\Tsys}{\emm{T_\emr{sys}}}
\newcommand{\Tas}{\emm{T_\emr{A}^*}}
\newcommand{\Tmb}{\emm{T_\emr{mb}}}

\shorttitle{Spatially resolved $l$-C$_3$H$^+$ emission in the Horsehead PDR}
\shortauthors{V.V. Guzm\'an et al.}

\begin{document}


\title{Spatially resolved \lowercase{$l$}-C$_3$H$^+$ emission in the
  Horsehead photodissociation region:\\ Further evidence for a
  top-down hydrocarbon chemistry\footnotemark[$\star$]}

\footnotetext[$\star$]{Based on observations obtained with the IRAM Plateau de
  Bure interferometer and 30~m telescope. IRAM is supported by
  INSU/CNRS (France), MPG (Germany), and IGN (Spain).}


\author{V.V. Guzm\'an\altaffilmark{1}}
\email{vguzman@cfa.harvard.edu}

\author{J. Pety\altaffilmark{2,3}}

\author{J.R. Goicoechea\altaffilmark{5}}

\author{M. Gerin\altaffilmark{3,4}}

\author{E. Roueff\altaffilmark{6,4}}

\author{P. Gratier\altaffilmark{7,8}}

\author{K.I. \"Oberg\altaffilmark{1}}

\altaffiltext{1}{Harvard-Smithsonian Center for Astrophysics, 60 Garden Street, Cambridge, MA 02138, USA}
\altaffiltext{2}{Institut de Radioastronomie Millim\'etrique (IRAM), 300 rue de la Piscine, 38406 Saint Martin d'H\`eres, France}
\altaffiltext{3}{LERMA, Observatoire de Paris, \'Ecole Normale Sup\'erieure, PSL Research
University, CNRS, UMR8112, F-75014, Paris, France}
\altaffiltext{4}{Sorbonne Universit\'es, UPMC Univ. Paris 06, UMR8112, LERMA, F-75005, Paris, France}
\altaffiltext{5}{Instituto de Ciencia de Materiales de Madrid (CSIC). E-28049 Cantoblanco, Madrid, Spain}
\altaffiltext{6}{LERMA, Observatoire de Paris, PSL Research University, CNRS, UMR8112, F-92190 Meudon, France}
\altaffiltext{7}{Univ. Bordeaux, LAB, UMR 5804, F-33270, Floirac, France}
\altaffiltext{8}{CNRS, LAB, UMR 5804, F-33270, Floirac, France}

\begin{abstract}
 Small hydrocarbons, such as \cch{}, \ccch{} and \ccchh{} are more
 abundant in photo-dissociation regions (PDRs) than expected based on
 gas-phase chemical models. To explore the hydrocarbon chemistry
 further, we observed a key intermediate species, the hydrocarbon ion
 \lccchp{}, in the Horsehead PDR with the Plateau de Bure
 Interferometer at high-angular resolution ($6''$). We compare with
 previous observations of \cch{} and \Cccchh{} at similar angular
 resolution and new gas-phase chemical model predictions to constrain
 the dominant formation mechanisms of small hydrocarbons in low-UV
 flux PDRs. We find that, at the peak of the HCO emission (PDR
 position), the measured \lccchp{}, \cch{} and \Cccchh{} abundances
 are consistent with current gas-phase model predictions. However, in
 the first PDR layers, at the 7.7~$\mu$m PAH band emission peak, which
 are more exposed to the radiation field and where the density is
 lower, the \cch{} and \Cccchh{} abundances are underestimated by an
 order of magnitude. At this position, the \lccchp{} abundance is also
 underpredicted by the model but only by a factor of a few. In
 addition, contrary to the model predictions, \lccchp{} peaks further
 out in the PDR than the other hydrocarbons, \cch{} and
 \Cccchh{}. This cannot be explained by an excitation effect. Current
 gas-phase photochemical models thus cannot explain the observed
 abundances of hydrocarbons, in particular in the first PDR
 layers. Our observations are consistent with a top-down hydrocarbon
 chemistry, in which large polyatomic molecules or small carbonaceous
 grains are photo-destroyed into smaller hydrocarbon
 molecules/precursors.
\end{abstract}        






\newcommand{\TabObs}{%
  \begin{table*}
    \begin{center}
    {\small \caption{Observation parameters for the maps. Their
        projection center is $\alpha_{2000} = 05^h40^m54.27^s$,
        $\delta_{2000} = -02\deg 28' 00''$.}} 
    \label{tab:obs:maps}
       \begin{threeparttable}
        \begin{tabular}{lrcccccccccr}\toprule
          Line & Frequency & $E_u/k$ & Instrument & Beam & PA & Vel. Resol. & Int. Time$^b$ & \Tsys{} & Noise \\
          & GHz & K & & arcsec & $^{\deg}$ & \kms{} & hours & K (\Tas{}) & K (\Tmb{})\\ 
          \midrule
          \Cccchh{} $2_{2,1}-1_{0,1}$ & 85.339 & 6.4 & PdBI/C\&D & $6.1\times4.7$ & 36 & 0.2 & 12.0 & -- & 0.30\\          
          \cch{} $N=1-0^a$ & 87.317 & 4.2 & PdBI/C\&D & $7.2\times5.0$ & 54 & 0.2 & 12.0 & -- & 0.60\\
          \lccchp{} $J=5-4$ & 112.446 & 16.2 & PdBI/C\&D & $6.2 \times 5.5$ & 25 & 0.2 & 17.3 & 150 & 0.07 \\
          \dcop{} $J=2-1$ & 144.077 & 10.4 & 30m/CD150 & $18.0\times18.0$ & 0 & 0.1 & 1.5 & 230 & 0.10\\ 
          \bottomrule
        \end{tabular}
             \begin{tablenotes}[para,flushleft]
               $^a$ Transition: $N=1-0, J=3/2-1/2, F=2-1$. $^b$ On-source integration time scaled to a 6 antenna array.
             \end{tablenotes}
       \end{threeparttable}
    \end{center}
\end{table*}
}

\newcommand{\TabColumnDensities}{%
  \begin{table*}
    \begin{center}
      {\small \caption{Column densities and abundances with respect to H nuclei.}} 
      \begin{tabular}{lccc}\toprule
        & CORE & PDR & PAH \\
        \midrule
        \NH{}         & $6.4\times10^{22} \pccm$ & $3.8\times10^{22} \pccm$ & $6.4\times10^{21} \pccm$\\
        \midrule
        $N$(\cch{})      & $<8.8\times10^{13}$  & $(1.3-1.6)\times10^{14}$ & $(1.5-5.0)\times10^{14}$\\
        $N$(\Cccch{})    & $(0.8-2.3)\times10^{12}$  & $(2.4-7.2)\times10^{12}$ & -                       \\
        $N$(\Lccch{})    & $(1.4-4.2)\times10^{11}$  & $(0.6-1.8)\times10^{12}$ & -                       \\
        $N$(o-\Cccchh{}) & $(2.3-3.0)\times10^{12}$  & $(3.8-6.9)\times10^{12}$ & $(1.2-2.7)\times10^{13}$\\
        $N$(p-\Cccchh{}) & $(0.6-1.1)\times10^{12}$  & $(1.2-1.9)\times10^{12}$ & -                       \\
        $N$(\Lccchh{})   & $(0.9-2.6)\times10^{11}$  & $(0.5-1.5)\times10^{12}$ & -                       \\
        $N$(\lccchp{})   & $<6.5\times10^{10}$       & $(1.6-4.8)\times10^{11}$ & $(1.4-4.1)\times10^{11}$\\
        \midrule                                                                
        X(\cch{})   & $<1.8\times10^{-9}$  & $(1.9-5.9)\times10^{-9}$ & $(1.3-9.0)\times10^{-8}$\\
        X(\ccch{})  & $(0.9-5.1)\times10^{-11}$ & $(0.5-2.7)\times10^{-10}$ & -\\
        X(\ccchh{}) & $(2.8-8.8)\times10^{-11}$ & $(0.9-3.2)\times10^{-10}$ & $(1.5-6.6)\times10^{-9}$\\
        X(\lccchp{}) & $<1.7\times10^{-12}$ & $(0.2-1.4)\times10^{-11}$ & $(1.2-7.2)\times10^{-11}$\\
        \bottomrule 
      \end{tabular}
      \label{tab:columns}
    \end{center}
  \end{table*}
}

\newcommand{\FigMaps}{%
\begin{figure*}
  \begin{center}
  \includegraphics[width=0.8\textwidth]{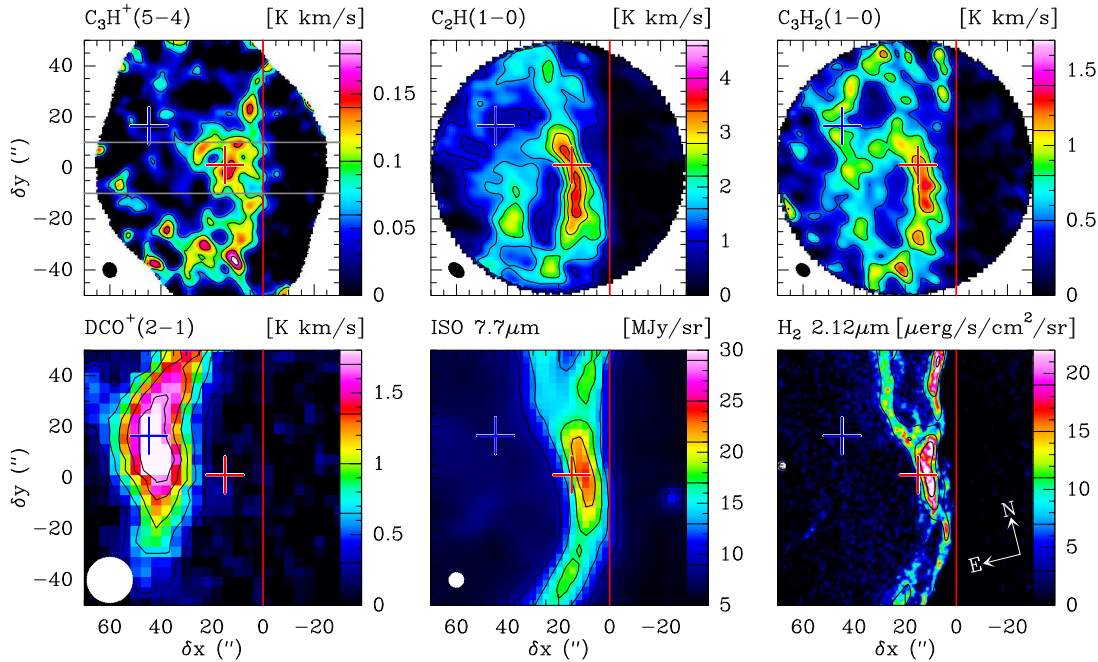}
  \caption{Integrated intensity maps of the small hydrocarbons
    \lccchp{}, \cch{} and \Cccchh{} lines (upper row), as well as that of
    the \dcop{} line, the 7.7 $\mu$m PAH emission and the \hh{} 2.12 $\mu$m
    ro-vibrational line (bottom row). Maps have been rotated by
    14$^{\deg}$ counter-clockwise around the projection center,
    located at $(\delta x,\delta y) = (20'',0'')$, to bring the
    illuminating star direction in the horizontal direction and the
    horizontal zero has been set at the PDR edge, delineated by the
    red vertical line. The blue and red crosses show the Core and PDR
    positions, respectively. The two gray horizontal lines on top of
    the \lccchp{} map display the region over which the spectra are
    averaged in Fig.~\ref{fig:cut}. All maps have been integrated
    between 10.1 and 11.1 $\kms$.}
  \label{fig:maps}
  \end{center}
\end{figure*}
}

\newcommand{\FigCuts}{%
\begin{figure}
  \begin{center}
  \includegraphics[width=0.47\textwidth]{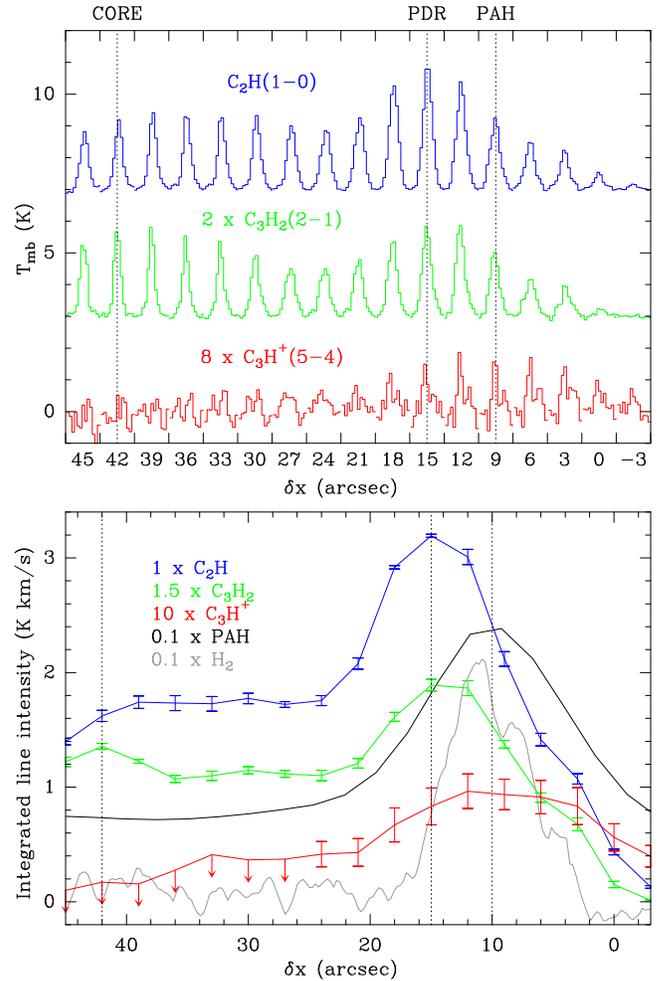}
  \caption{\textit{Left:} Spectra along the direction of the exciting
    star at the PDR position ($dy=0$ in Fig.~\ref{fig:maps}), averaged
    over $20''$ in the $dy$-direction. \textit{Right:} Integrated line
    intensities along the direction of the exciting star at the PDR
    position.}
  \label{fig:cut}
  \end{center}
\end{figure}
}

\newcommand{\FigModel}{%
\begin{figure}[t!]
  \begin{center}
  \includegraphics[width=0.45\textwidth]{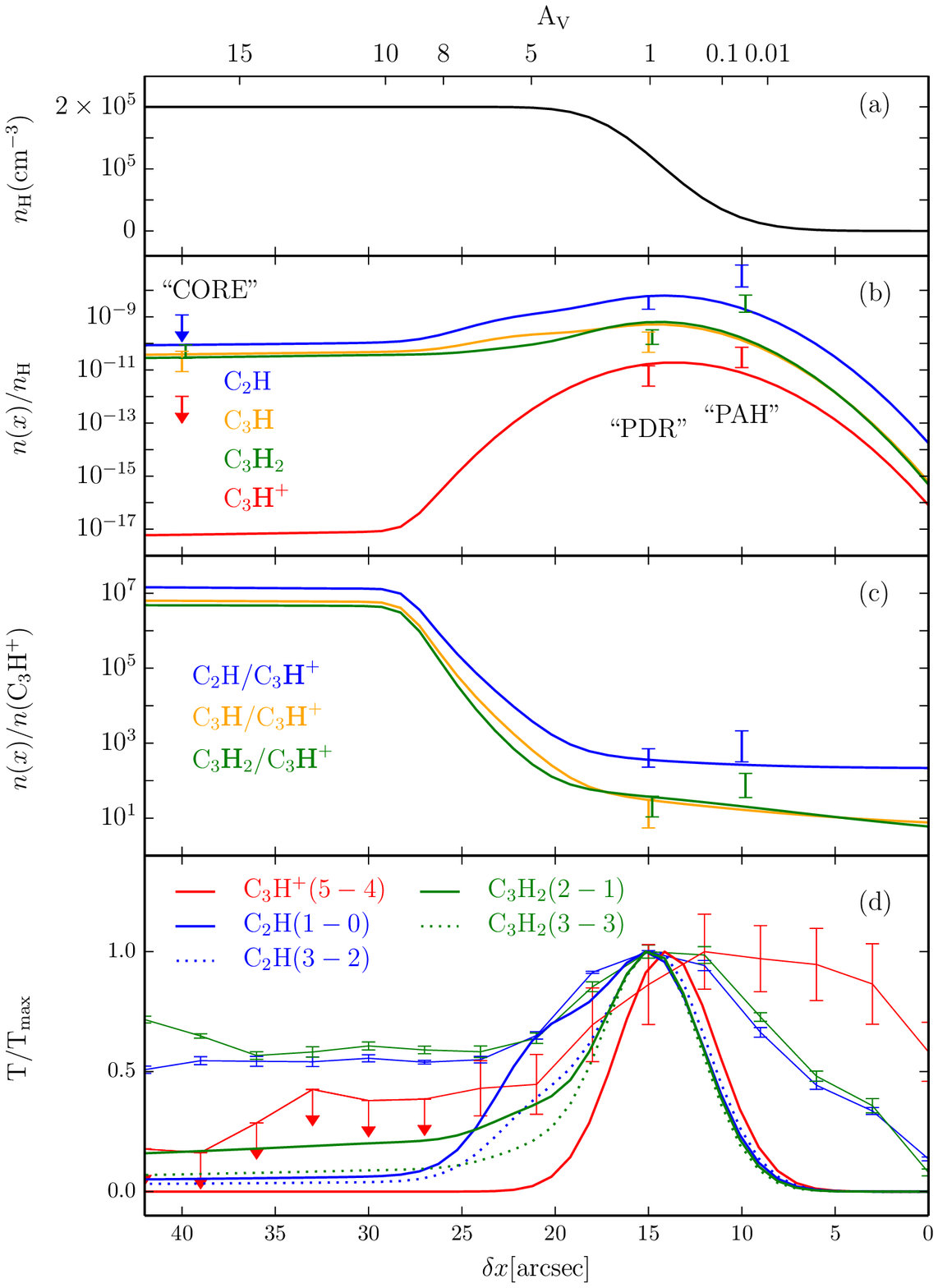}
  \caption{Photochemical model of the Horsehead PDR. (a) Density
    profile $n_{\h} = n(\h) + 2 n(\hh)$. (b) Predicted abundance of
    small hydrocarbons. (c) Predicted abundance ratio relative to
    \lccchp{}. (d) Predicted intensity profile. Two lines are included
    for \cch{}: $N=1-0$ ($E_u/k\sim$4~K) and $N=3-2$
    ($E_u/k\sim25$~K); and for \Cccchh{}: $J=2-1$ ($E_u/k\sim6$~K) and
    $J=3-2$ ($E_u/k\sim16$~K). The observed intensity profile is
    overlaid with error bars.}
  \label{fig:model}
  \end{center}
\end{figure}
}


\section{Introduction}

\TabObs{}
\FigMaps{}

Simple hydrocarbon molecules, such as \cch{}, \ccch{}, \ccchh{} and
\cccch{} are ubiquitous in the interstellar medium (ISM). They are
easily observed in a large variety of sources, from diffuse
\citep[\eg{}][]{lucas2000} to dark clouds
\citep[\eg{}][]{wootten1980,cox1989,mangum1990}. In Photon Dominated
Regions (PDRs) they have been found to be almost as abundant as in
dark, well-shielded clouds, despite the strong UV radiation compared
to the mean interstellar radiation field
\citep[\eg{},][]{fosse2000,fuente2003,teyssier2004,pety2005b}. In
contrast to high UV-flux PDRs \citep[$\chi\sim10^4-10^5$ relative to
  the Draine field;][]{draine1978}, like Mon~R2 or the Orion Bar,
where the observed hydrocarbon abundances can roughly be explained
with pure gas-phase chemistry models \citep{cuadrado2014}, in low-UV
flux PDRs ($\chi\lesssim100$) the current pure gas-phase chemical
models fail to reproduce their high abundances. \citet{teyssier2004}
and \citet{pety2005b} proposed that another mechanism producing carbon
chains must exist in addition to gas-phase chemistry.  One such
possibility, suggested by some laboratory experiments and theoretical
calculations, is the fragmentation of polycyclic aromatic hydrocarbons
(PAH) or very small carbonaceous grains (VSGs) due to the far-UV
radiation field \citep[][and references
  therein]{lepage2003,montillaud2013}. Indeed, recent laboratory
experiments have shown that the far-UV irradiation of interstellar
hydrogenated amorphous carbon analogs leads to the efficient
production of small hydrocarbons, such as CH$_4$
\citep{alata2014}. The good spatial correlation between the mid-IR
emission due to PAHs and the distribution of carbon chains found in
the Horsehead nebula, a low-UV flux PDR ($\chi\sim60$), provides some
support for a hydrocarbon production mechanism starting from PAHs
\citep{pety2005b}.

Using single-dish observations, \citet{pety2012} reported the first
detection of \lccchp{}, a key intermediate species in the gas-phase
formation of small hydrocarbons, toward the Horsehead PDR. \lccchp{}
was later detected toward the Sgr~B2(N) molecular cloud
\citep{mcguire2013}. \citet{brunken2014} measured the millimeter
rotational spectrum of \lccchp{} in the laboratory and unambiguously
confirmed the assignment of the observed lines in the Horsehead to the
hydrocarbon ion, \lccchp{}. Moreover, \citet{botschwina2014} performed
highly accurate quantum chemical calculations to study this ion and
found an excellent agreement with the spectroscopic constants derived
from the observations in the Horsehead. More recently,
\citet{cuadrado2014} detected the \lccchp{} lines up to $J=13$ in the
Orion Bar, what allowed them to further refine the rotational
constants. In addition, \cite{mladenovic2014} revised the
spectroscopic parameters of \lccchp{} by means of numerically exact
rovibrational calculations. The detection of the small hydrocarbon
\lccchp{} has thus been confirmed, both theoretically and
experimentally.

Chemically, \ccchp{} is a gas-phase precursor of the small
hydrocarbons \ccch{} and \ccchh{}. Observations of the distribution of
\ccchp{} with respect to these hydrocarbons could thus be used to
constrain the dominant formation mechanism of small hydrocarbons. The
Horsehead Nebula provides an ideal test-bed because it is viewed
almost edge-on, providing easy access to the warm surface layer of a
cloud where \cp{} and therefore \ccchp{} is predicted to be the most
abundant. As this warm photo-active layer is spatially narrow
\citep[$\sim5''$,][]{guzman2012}, high angular resolution is needed to
resolve the steep gradients in this region. To explore the
relationship of \ccchp{} with its environment and with neutral carbon
chains, we present in this letter the first spatially resolved
observations of \lccchp{}.

\section{Observations}

We used the Plateau de Bure Interferometer (PdBI) to obtain a $6''$
angular resolution map of the emission of the \lccchp $(5-4)$ line at
112.446~GHz. The observation parameters are summarized in
Table~\ref{tab:obs:maps}. The observations were carried out in
December 2012, April, May, September and October 2013 with six
antennas in the C and D configurations (baseline lengths between 24
and 176~m). We observed a 9-field mosaic and used about 37 hours of
telescope time, which correspond to 17.3 hours of on-source time
scaled to a six antenna array after filtering out low-quality
visibilities. The \lccchp{} line was covered with a correlator window
of 20~MHz bandwidth and 39~kHz channel spacing. The typical
precipitable water vapor amounted to 6~mm and the typical system
temperature was 150~K.

The PdBI data was calibrated with the \GILDAS{}\footnote{See
  \texttt{http://www.iram.fr/IRAMFR/GILDAS} for more information about
  the \GILDAS{} softwares~\citep{pety2005a}.}/\CLIC{} software. The
bright quasars 3C84, 2200$+$420 and 3C279 were used to calibrate the
radio-frequency bandpass, and two nearby quasars (0420$-$014 and
0528$+$134) were regularly observed to calibrate phase and amplitude
temporal variations. MWC349 was used to derive the absolute flux
scale.  In order to recover the extended emission that is filtered out
by the PdBI, we observed the same region with the IRAM-30m telescope
during $\sim15$ hours of average summer weather in July and October
2013.  The \GILDAS{}/\CLASS{} software was used to process the
IRAM-30m data and produce the single-dish map, which was then combined
with the PdBI observations in \GILDAS{}/\MAPPING{}, in the same way as
described in \citet{guzman2013}. 

\section{Results}   

\subsection{Spatial distribution}

Fig.~\ref{fig:maps} displays the integrated intensity maps, at similar
angular resolution ($6''$), of the \lccchp{} $J=5-4$, \cch{}
$N=1-0,J=3/2-1/2,F=2-1$ and \Cccchh{} $2_{2,1}-1_{0,1}$ lines
\citep{pety2005b}. Also displayed are the 7.7~$\mu$m emission arising
from PAHs imaged with ISO at $6''$ angular resolution
\citep{abergel2003}, and the integrated intensity maps of the \dcop{}
$J=2-1$ line observed with the IRAM-30m telescope \citep{pety2007} and
the \hh{} $v=1-0$ $S(1)$ line obtained with SOFI/NTT at $1''$ angular
resolution \citep{habart2005}. 

\FigCuts{}

The emission of the \cch{} and \Cccchh{} lines is structured into two
successive filaments parallel to the dissociation front. The first one
mainly coincides with the PDR, where intense emission is seen both in
the \hh{} ro-vibrational line and in the PAH mid-infrared band. The
second one is located at the position of the cold dense UV-shielded
core as indicated by the bright \dcop{} emission, with a difference
between the two species: At the peak of the \dcop{} emission there is
a clear, localized deficit of the \cch{} emission compared to that of
\Cccchh{}. In contrast, the \lccchp{} $J=5-4$ line only emits toward
the UV-illuminated edge, at the surface of the cloud. Moreover, the
\lccchp{} emission reaches the red line in Fig.~\ref{fig:maps}, which
traces the edge of the PDR, while the emission of the two other
hydrocarbons is shifted left of this line. We note that the observed
\lccchp{} line is weak ($6\sigma$ at the PDR), and hence the clumpy
structure is most likely an artifact caused by the low signal-to-noise
ratio. Indeed, due to the low inclination of the source the dirty beam
has side-lobes which produce some uncertainty in the deconvolution of
the data.


Line spectra along the direction of the illuminating star, centered at
$dy=0''$, are shown in the upper panel of Fig.~\ref{fig:cut}. To
increase the signal-to-noise ratios we have averaged the spectra over
$20''$ in the $dy$-direction, \ie{}, the region between the two gray
lines in Fig.~\ref{fig:maps}. In the bottom panel the integrated line
intensity profiles of the hydrocarbons are shown, as well as that of
the \hh{} line and the PAH emission. The dashed vertical lines mark
three characteristic positions in the Horsehead: the Core,
corresponding to the peak of the \dcop{} line emission (RA
$=5^\mathrm{h}40^\mathrm{m}55.61^\mathrm{s}$, Dec $=-2\deg27'38''$,
J2000), and characteristic of the cold UV-shielded gas (marked by the
blue cross in Fig~\ref{fig:maps}); the PDR, corresponding to the peak
of the HCO line emission emission \citep[RA
  $=5^\mathrm{h}40^\mathrm{m}53.936^\mathrm{s}$, Dec $=-2\deg28'00''$,
  J2000;][]{gerin2009} and characteristic of the warmer UV-illuminated
gas (marked by the red cross in Fig.~\ref{fig:maps}); and a position
closer to the edge of the cloud, corresponding to the peak of the
7.7~$\mu$m PAH emission (RA
$=5^\mathrm{h}40^\mathrm{m}53.6^\mathrm{s}$, Dec= $-2\deg28'1.9''$,
J2000). We observe a shift between the peak of \lccchp{} emission
compared to that of the other hydrocarbons. \lccchp{} peaks at $\delta
x \sim 10''$, while the other hydrocarbons peak at $\delta x \sim
15''$. Although the peak position of the \lccchp{} line has a higher
uncertainty given the lower signal-to-noise ratio, the emission
profile is clearly broader and it extends further out in the PDR
compared to the \cch{} and \Cccch{}. Moreover, the \lccchp{}
integrated line intensity profile correlates better to that of the
\hh{} line and the PAHs emission compared to neutral hydrocarbons.

\subsection{Abundances}
\label{sec:abundances}
\TabColumnDensities{}

To compare with chemical models, we have computed the column densities
of \cch{}, \ccch{}, \ccchh{} (both linear and cyclic species) and
\lccchp{} at three different positions, namely the Core, PDR and PAH
positions. These positions are slightly different from those used by
\citet{pety2005b} and \citet{pety2012}, and were chosen to take
advantage of Horsehead WHISPER line survey. For the first two
positions, we used the single-dish deep integrations obtained in the
line survey and included beam dilution factors obtained from the
higher-angular resolution PdBI observations when available. As no line
survey has been made at the PAH position, we derived the abundances
directly from the PdBI observations. For \ccch{}, we assumed the same
spatial distribution of \Cccchh{}. We used the non-LTE radiative
transfer code RADEX \citep{vandertak2007} for those species with known
collisional rates, \ie{}, \cch{} and \Cccchh{} (ortho and para), taken
from \citet{spielfiedel2012} and \citet{chandra2000},
respectively. The gas density was fixed to $6\times10^4 \pccm{}$ in
the PDR, $10^5 \pccm{}$ in the dense core \citep{gerin2009,pety2007}
and $(5-10)\times10^3 \pccm$ in the PAH position. The kinetic
temperature was left as a free parameter, the best fits being
consistent with previous estimates of $\sim60$~K and $\sim20$~K in the
PDR and dense core, respectively. For \ccch{} and \Lccchh{} we
constructed rotational diagrams. We note that \cch{} and \lccchp{} are
not detected at the core position in the PdBI maps. We thus consider
their derived abundances as upper limits.  The inferred column
densities and abundances with respect to total hydrogen nuclei are
summarized in Table~2. The errors in the abundances take into account
a 50\% uncertainty in the assumed \NH{}, which are inferred from the
1.2~mm dust continuum emission.

\citet{pety2012} derived the \lccchp{} abundance at the PDR position
using single-dish observations at $\sim25''$ angular resolution, and
computing a beam filling factors by assuming that the \lccchp{}
emission filled a Gaussian filament of $\sim12''$ width in the $\delta
x$ direction. This assumption was based on the morphology of the HCO
line emission, which shows a filament of roughly this width. The new
PdBI observations at 6$''$ angular resolution show that the \lccchp{}
emission indeed arises from a $\sim12''$ filament, where the
\lccchp{}~$J=5-4$ line is $\sim3$ times brighter than what was
observed with the 30m.

\subsection{Chemistry}   
\label{sec:chem}

In order to test our current knowledge of the gas-phase chemistry of
hydrocarbons, we have used an updated version of the one-dimensional,
steady-state photochemical code from \citet{lepetit2006}. The same
model was used in \citet{pety2012}, except we have now introduced the
formation and destruction (by \hh{}) rates of \ccchp{} measured by
\cite{savic2005}, which have an inverse temperature dependence. For
the physical conditions in the Horsehead ($\nH\simeq6\times10^4\pccm$,
$\Tkin\sim60$~K), this results in higher abundances of \ccch{} and
\ccchh{}. The model includes grain surface reactions for the formation
of \hh{} and other species, such as H$_2$CO and CH$_3$OH, but only
gas-phase reactions for the formation of hydrocarbons \citep[see][ for a detailed description of the
  model]{pety2012}.

In high UV-flux PDRs ($\Tkin\sim100-500$~K), the formation of
hydrocarbons starts with the formation of CH$^+$ \citep{cuadrado2014}
through the very endothermic reaction $\cp + \hh \rightarrow$ CH$^+ +$
H \citep{agundez2010}. In the Horsehead PDR, on the other hand, we
find that the hydrocarbon gas-phase chemistry is initiated by
reactions of \cp{} with CH leading to C$^+_2$. Further reactions with
\hh{} lead to the formation of C$_2$H$^+$, C$_2$H$^+_2$ and
C$_2$H$^+_3$. The last two species recombine with electrons to form
\cch{} and C$_2$H$_2$, respectively. C$_2$H$_2$ can react with \cp{}
to form \ccchp{}, but in fact the dominant formation route of \ccchp{}
at the edge of the cloud involves reactions between \cch{} and \cp{}
leading to C$_3^+$, followed by reactions with \hh{} in the model.
Once \ccchp{} is produced, it reacts with \hh{} to form the
C$_3$H$^+_2$ and C$_3$H$^+_3$ ions which then recombine with electrons
to form \ccch{} and \ccchh{}. \ccchp{} can also recombine with
electrons to form \cch{}, although the dominant formation route for
\cch{} is the recombination of C$_2$H$^+_2$ with electrons:
\begin{equation*} 
  {\small
  \xymatrixrowsep{0.5cm}
  \xymatrixcolsep{0.5cm}
  \xymatrix{
   \chem{C_2H^+} \ar[dd]_{\tiny \chem{H_2}} & \chem{C_2^+} \ar[l]_{\tiny \chem{H_2}}  & \chem{CH} \ar[l]_{\tiny \chem{C^+}} & & & \\
   & \chem{C_2H_3^+} \ar[r]^{\tiny \chem{e^{-}}}  & \chem{C_2H_2} \ar[rd]^{\tiny \chem{C^+}} & & \chem{C_3H^+_3} \ar[r]^{\tiny \chem{e^{-}}}  & \color{green}{\chem{C_3H_2}} \\
   \chem{C_2H_2^+}  \ar[ru]^{\tiny \chem{H_2}} \ar[rd]_{\tiny \chem{e^{-}}}  & & & \color{red}{\chem{C_3H^+}}  \ar[lld]_{\tiny \chem{e^{-}}} \ar@{.>}[ru]^{\tiny \chem{H_2}} \ar@[gray][rd]_{\tiny \chem{H_2}}  \ar[r]^{\tiny \chem{e^{-}}} & \chem{C_3} & \\
   & \color{blue}{\chem{C_2H}} \ar[r]_{\tiny \chem{C^{+}}} & \chem{C_3^+} \ar@[gray][ru]_{\tiny \chem{H_2}} & &  \chem{C_3H^+_2} \ar[r]^{\tiny \chem{e^{-}}}  & \color{orange}{\chem{C_3H}} 
    }}
\end{equation*}
The dotted arrow in the scheme above marks a radiative association
reaction, and the gray arrows indicate reactions with a temperature
dependence of the rates. The \ccchp{} hydrocarbon ion is thus a key
chemical precursor of the small hydrocarbons in the gas phase. In
particular, one would expect \ccchp{} and \ccchh{} to have a similar
spatial distribution if gas-phase chemistry alone is responsible for
the observed \ccchh{} abundance.

Fig.~\ref{fig:model} displays the results of the photochemical model
convolved to a resolution of $6''$ to facilitate the comparison with
observations. The steep density profile of the Horsehead is shown in
the upper panel (a). The abundances with respect to total hydrogen
nuclei are shown in panel (b). The abundance ratios with respect to
\lccchp{} are shown in the panel (c). The observations are shown with
error bars.  In general, the abundances are well-reproduced by the
pure gas-phase model at the PDR position ($A_V\simeq1.5$). At the PAH
position ($A_V\simeq0.05$), on the other hand, the \cch{} and
\Cccchh{} abundances are underpredicted by an order of magnitude. The
same conclusion can be drawn from the abundance ratios, where the
match between observations and model is better at the PDR than at the
PAH position. We note that at the PAH position, the model better
reproduces the observed \lccchp{} abundance than that of \cch{} and
\ccchh{}. This suggests the need of an additional formation mechanism
to explain the observed hydrocarbon abundances at the edge of the
cloud.

\FigModel{}

The bottom panel (d) in Fig.~\ref{fig:model} shows the modeled line
intensity profile for each species and for different rotational
transitions, normalized to their respective emission peaks. The
observed profiles are overlaid with error bars. The modeled line
intensity profiles of \cch{} and \Cccchh{} were computed by including
the PDR model outputs (\hh{} density, $\Tkin$ and column density) into
the non-LTE radiative transfer code RADEX. In order to investigate
excitation effects on the position of the emission peak, we included
two transitions with different upper level energies for each species:
the \Cccchh{}~$2_{1,2}-1_{0,1}$ ($E_u=6$~K) and \cch{}~$N=1-0,
J=3/2-1/2, F=2-1$ ($E_u=4$~K) (solid lines), and the
\Cccchh{}~$3_{1,2}-3_{0,3}$ ($E_u=16$~K) and $\cch~N=3-2, J=7/2-5/2,
F=3-2$ ($E_u=25$~K) (dashed lines). These lines were chosen
because their upper level energies are closer to that of the \lccchp{}
$J=5-4$ line ($E_u/k\sim16$~K). Because there are no available
collisional coefficients for \lccchp{}, we assumed a constant
excitation temperature along the cloud to compute its line intensity
profile. In the model all hydrocarbons peak at the same position
($\delta x \sim 15''$). However, our observations show that the
\lccchp{} emission peak is shifted compared to that of the other
hydrocarbons. We note that the emission peak of \cch{} and \Cccchh{}
does not change significantly between the lower and higher energy
transitions. We have checked that including a constant excitation
temperature for \cch{} and \Cccch{} gives similar results. This
suggests that the excitation temperature gradient in the Horsehead
takes place at very small spatial scales, and that the observed shift
in the \lccchp{} emission peak compared to the other hydrocarbons is
not due to an excitation effect but reveals a real difference in the
spatial distribution of their column densities that is not predicted
by the current pure gas-phase chemical models.

In general, the observed integrated intensity profile of all species
is much broader than what the model predicts. In particular, the model
underpredicts the \lccchp{} emission in the first PDR layers ($\delta
x <10''$). We run models with constant densities ($10^4-10^5 \pccm$)
to explore the effect of the density profile on the \lccchp{}
intensity in the surface layers. We find that a higher density would
slightly improve the agreement between model and observations, but it
would not reproduce the observations of many other tracers (\eg{},
HCO, \hh{}, and dust continuum emission). Therefore, an additional
formation mechanisms such as PAH photodestruction contributes to the
formation of \lccchp{} in these surface layers. Indeed, it has been
suggested that C$_2$H$_2$, which would enhance the \ccchp{} abundance
through reactions with \cp{}, could be a product of photochemistry of
PAHs \citep{bierbaum2011}.

\section{Conclusions}

We have presented high-angular resolution ($6''$) observations of the
hydrocarbon ion \lccchp{} in the Horsehead photo-dissociation
region. The \lccchp{} emission is concentrated toward the surface edge
of the nebula, close to where other hydrocarbon chains, such as \cch{}
and \Cccchh{}, and the 7.7~$\mu$m PAH emission peak. However, in
contrast to \cch{} and \Cccchh{}, \lccchp{} is only abundant in the
first PDR layers with almost no emission deeper inside the
cloud. \lccchp{} is thus a good tracer of hydrocarbon
photo-chemistry. Moreover, \lccchp{} peaks $\sim5''$ further outside
the PDR than \cch{} and \Cccchh{}, which cannot be explained by an
excitation effect. This is in contrast to what current gas-phase
chemical models predict for the Horsehead physical conditions, that
is, that all the hydrocarbons should peak approximately at the same
position as their chemistry is closely linked.

The inferred \lccchp{} abundance at the PDR from the new
high-resolution observations is consistent with that derived by
\citet{pety2012}. In addition, we have computed the abundances of
\cch{}, \ccchh{} and \lccchp{} at three characteristic positions in
the cloud. We find that a gas-phase chemical model can reproduce the
hydrocarbon abundances, including the \lccchp{} abundance, at the PDR
but it underpredicts the \cch{}, \Cccchh{} and \lccchp{} abundances at
the PAH position, \ie{}, at the edge of the cloud or the first PDR
layers. The fact that the disagreement between model and observations
occurs at the edge of the cloud, where the PAHs emission peaks, is
consistent with a top-down hydrocarbon chemistry, where PAHs and small
carbonaceus grains are photo-eroded by the radiation field releasing
small hydrocarbons into the gas-phase. The Horsehead PDR is a clear
case where this mechanism is efficient. Laboratory experiments are,
however, needed to quantify the amount of hydrocarbons and the
specific products that can be produced from this mechanism, and how
this affects the hydrocarbon chemistry.

\acknowledgments We thank the IRAM PdBI and 30~m staff for their
support during the observations. This work was partially funded by the
CNRS Programme Nationale de Physique et Chimie du Milieu
Interstellaire (PCMI).
J.R.G. thanks the Spanish MINECO for funding
  support from grants CSD2009-00038, AYA2009-07304 and AYA2012-32032.
PG acknowledges funding from the ERC Starting Grant 3DICE (336474).

 {\it Facilities:} \facility{IRAM:Interferometer}, \facility{IRAM:30m}.

\end{document}